\documentclass[sigconf]{acmart}
\usepackage{makecell, url}
\newcommand{\highlightcolor}{black}
\newcommand{\colortext}[1]{\textcolor{\highlightcolor}{#1}}
\AtBeginDocument{%
  }

\setcopyright{acmlicensed}
\copyrightyear{2026}
\acmYear{2026}
\acmDOI{XXXXXXX.XXXXXXX}
\acmConference[Conference acronym 'XX]{Make sure to enter the correct
  conference title from your rights confirmation email}{April 13--17,
  2026}{Barcelona, Spain}
\copyrightyear{2026}
\acmYear{2026}
\setcopyright{cc}
\setcctype{by}
\acmConference[CHI '26]{Proceedings of the 2026 CHI Conference on Human Factors in Computing Systems}{April 13--17, 2026}{Barcelona, Spain}
\acmBooktitle{Proceedings of the 2026 CHI Conference on Human Factors in Computing Systems (CHI '26), April 13--17, 2026, Barcelona, Spain}
\acmPrice{}
\acmDOI{10.1145/3772318.3791647}
\acmISBN{979-8-4007-2278-3/2026/04}

\begin{document}

\title[Metacognitive Demands of AI Conversational Agents for Health Information]{Metacognitive Demands and Strategies While Using Off-The-Shelf AI Conversational Agents for Health Information Seeking}

\author{Shri Harini Ramesh}
\orcid{0009-0002-4245-7693}
\affiliation{%
  \institution{University of Calgary}
  \city{Calgary}
  \state{Alberta}
  \country{Canada}
}
\email{shri.ramesh@ucalgary.ca}

\author{Foroozan Daneshzand}
\affiliation{%
  \institution{Simon Fraser University}
  \city{Burnaby}
  \state{British Columbia}
  \country{Canada}}
\email{foroozan_daneshzand@sfu.ca}

\author{Babak Rashidi}
\affiliation{%
  \institution{Ottawa General Campus}
  \city{Ottawa}
  \state{Ontario}
  \country{Canada}}
\email{brashidi@toh.ca}

\author{Shriti Raj}
\affiliation{%
  \institution{Stanford University}
  \city{Palo Alto}
  \state{California}
  \country{United States}}
  \email{shritir@stanford.edu}

\author{Hariharan Subramonyam}
\affiliation{%
  \institution{Stanford University}
  \city{Palo Alto}
  \state{California}
  \country{United States}}
\email{harihars@stanford.edu}
\author{Fateme Rajabiyazdi}
\affiliation{%
  \institution{University of Calgary}
  \city{Calgary}
  \state{Alberta}
  \country{Canada}}
  \email{fateme.rajabiyazdi@ucalgary.ca}

\renewcommand{\shortauthors}{Ramesh et al.}

\begin{abstract}
  As Artificial Intelligence (AI) conversational agents become widespread, people are increasingly using them for health information seeking. The use of off-the-shelf conversational agents for health information seeking could place high metacognitive demands (the need for extensive monitoring and control of one’s own thought process) on individuals, which could compromise their experience of seeking health information. However, currently, the specific demands that arise while using conversational agents for health information seeking, and the strategies people use to cope with those demands, remain unknown.  To address these gaps, we conducted a think-aloud study with 15 participants as they sought health information using our off-the-shelf AI conversational agent. We identified the metacognitive demands such systems impose, the strategies people adopt in response, and propose considerations for designing beyond off-the-shelf interfaces to reduce these demands and support better user experiences and affordances in health information seeking.
\end{abstract}



\maketitle

\section{Introduction}
People regularly search for health information to interpret symptoms, explore possible causes, and treatment options~\cite{lila2019}. \colortext{They then appraise the credibility and relevance of the information they find and decide how to apply it to their personal health decisions, such as whether or not to seek professional care.}
Throughout this process, they use a variety of tools, including books, news articles, search engines~\cite{Choudhury2014}, health-specific websites such as WebMD~\cite{Zhao2017}, and patient communities hosted on social media platforms to search for health information~\cite{Presiado2024KFF}. 
However, as Artificial Intelligence (AI) conversational agents become widespread, people are increasingly turning to them for health information seeking~\cite{Yun2025healthinfo}. 
This may be because, instead of needing to go through multiple sources, these systems generate a single synthesized conversational response that appears tailored to a person’s specific question~\cite{Xiong2024searchEngine}. For many, this mode of engagement provides an immediate and comprehensive experience, reducing the effort of piecing together information across scattered sources~\cite{Sharma2024LLms}.

Despite these benefits, the use of AI conversational agents in health contexts raises some concerns. 
Health-related queries are particularly sensitive because inaccurate, incomplete, or misleading information can directly influence decisions about care~\cite{Zada2025misinfo, Wang2024misinfo}. 
While these conversational agents may simplify information gathering by providing consolidated responses, people \colortext{may still need digital health literacy~\cite{SMITH2019280} to engage in health information appraisal, evaluate whether the information is accurate, complete, and trustworthy, and consider its application to their own circumstances}~\cite{Yun2025trust}. 
However, people still face metacognitive demands when interacting with agents, which require them to plan, monitor, and regulate their own thinking as they interact with the agent~\cite{flavell1979metacognition}, regardless of being digitally literate~\cite{Tankelevitch2024}. 

At present, most people engage with these agents through off\-the‑shelf interfaces such as ChatGPT~\cite{openai_chatgpt2025}, Anthropic Claude~\cite{anthropic_claude2024}, and Google Gemini~\cite{google_bard2025}—these general-purpose interfaces are not specifically designed for health information seeking. 
While these interfaces are widely available and easy to use, they provide little guidance or scaffolding to help individuals manage the metacognitive demands of evaluating and acting on health information~\cite{Tankelevitch2024}. 
Thus, people are left to do this work on their own, often without realizing how much effort it actually requires. 
\colortext{Although previous research suggests that digital health literacy can support individuals' capacity to navigate and gather health information effectively~\cite{Manganello2017}, less attention has been paid to metacognitive skills, the critical awareness and self-regulation of one's own thinking and information practices, which are foundational to developing and applying digital health literacy in real-time interactions with AI agents~\cite{Lee2021MetaHealth}. Understanding how people seek health information with AI conversational agents through the lens of metacognition remains underexplored.}
This leaves a crucial gap in our understanding: \textit{how do metacognitive demands arise when people use off-the-shelf conversational agents for health information seeking, and what strategies do they adopt to cope with these demands?}

To address this gap, we conducted a think-aloud study with 15 participants to explore how people use AI conversational agents for health information seeking. In close collaboration with an internal medicine physician, we designed six common health scenarios that required people to interpret symptoms, weigh treatment options, and decide when to seek professional care. We randomly assigned two scenarios to each participant, asked them to imagine themselves in those situations and gather health information using the conversational agent we developed, powered by the ChatGPT-4o API. We built our agent interface to resemble current off-the-shelf AI conversational agent interfaces, while adding a custom backend to log and track user interactions for research purposes. 
We analyzed the study transcripts to capture the metacognitive demands people face and the strategies they employ when seeking health information. Based on these insights, we also propose design considerations for moving beyond off-the-shelf interfaces to reduce these metacognitive demands and support better user experiences in health information seeking. 
Our contributions are as follows:
\begin{itemize}
\item We examined the metacognitive demands that arise when people use AI conversational agents for health information seeking.
\item We identified the strategies people employed to monitor and regulate their interactions with AI conversational agents, highlighting when these strategies support effective information seeking and when they fall short.
\item Based on the identified metacognitive demands and strategies, we outlined design considerations for moving beyond off-the-shelf interfaces to reduce metacognitive demands and support better user experiences and affordances in health information seeking.
\end{itemize}

\section{Related Work}

Prior research has examined how people use different sources to seek health information to make decisions about their care. The emergence of AI conversational agents has introduced novel opportunities for accessing such health information, while simultaneously imposing metacognitive demands, as users must monitor their comprehension and determine when to trust or verify responses~\cite{Tankelevitch2024, lee2025impact}. Existing systems provide limited support for managing these challenges; thus, we need to design tools that can more effectively scaffold health information seeking~\cite{Zheng2024DiscipLink, Yan2025KNowNEt}. This section reviews prior work on health information seeking behaviors, the metacognitive demands associated with using conversational agents, and systems developed to assist people with these tasks.

\subsection{Health Information Seeking}
Over the past two decades, the way people seek health-related information has undergone a major transformation. 
Patients, their caregivers, and those proactively maintaining wellness get actively engaged in seeking health information, looking for information on their symptoms and diagnoses, as well as lifestyle advice such as diet and exercise~\cite{ghahramani2020impact}.
Traditionally, individuals relied on books and printed materials, but the internet quickly became the primary source for gathering information to address health questions~\cite{Hesse2005}. 
Some people turned to online health communities~\cite{luo2018exploring, lu2019impact} where they could connect with physicians, hospitals, and other medical ecosystems, and get access to information exchange platforms~\cite{hajli2014developing}. 
Beyond those, internet searches for health-related purposes are now common, with search engines such as Google serving as the most frequent starting point~\cite{bachl2024search}. Over the past two decades, research on health information seeking behavior has shown that people are increasingly turning to online resources~\cite{maon2017online,lee2014online} because of their availability~\cite{osei2017internet}, breadth of information ~\cite{lagoe2015health}, affordability~\cite{osei2017internet}, interactivity (e.g., participating in discussions or asking questions online)~\cite{powell2011characteristics}, and anonymity~\cite{jia2021online}. 

More recently, this shift has deepened with the emergence of  AI conversational agents, such as ChatGPT (OpenAI)~\cite{openai_chatgpt2025}, Bard (Google, now Gemini)~\cite{google_bard2025}, and Claude (Anthropic)~\cite{anthropic_claude2024}. These systems allow users to interact with AI through natural language conversations~\cite{Yun2025healthinfo} and draw from a wide range of online content, potentially offering users diverse perspectives they might not encounter when manually selecting sources~\cite{Chen2024}.
Unlike search engines that return a ranked list of links,  AI conversational agents provide direct, synthesized answers in a conversational style that feel immediate and coherent~\cite{Xiong2024searchEngine}. 
This conversational format reduces the need for users to evaluate and choose between multiple sources~\cite{Sharma2024LLms}.
Recent studies show that people perceive  AI conversational agents as more human-like, less biased, and eliciting fewer negative feelings~\cite{Mendel2025laypeople}, and that they elicit higher trust and satisfaction, particularly when responses include visible source citations~\cite{Yun2025trust}.
Together, these abilities position  AI conversational agents as a powerful tool for health information seeking—while also raising new questions about how users interpret and engage with the answers they receive~\cite{Yun2025trust}.

\subsection{Challenges and Metacognitive Demands with AI conversational agents}
Metacognition is the process of reflecting on and regulating one’s own thinking while learning or solving problems~\cite{flavell1979metacognition}.
It includes being aware of how one is learning, recognizing how problems are being solved, and taking active steps to guide or adjust those processes~\cite{Efklides2008}. 
For example, instead of jumping into a reading or learning task, a metacognitively aware person would first consider their task goals, how to achieve them, how well their current strategy is working, and whether they need to make changes along the way.
Although metacognitive abilities are essential for effective learning and problem-solving, being conscious and actively using these abilities is not always straightforward~\cite{Teng19102020}. 
People often do not monitor their understanding or evaluate their strategies unless they face difficulties or are prompted to do so, such as when a teacher or an instructor encourages them to reflect~\cite{SCHRAW1994, Wang1990}. 
Improving these skills or providing support that prompts learners to use them helps individuals recognize when their current approach isn’t working and make adjustments, leading to better learning and decision-making~\cite{Azevedo2010}.

Conversational agents need individuals to state their goals clearly using prompts, question the outputs, and decide when to trust or reject them. 
However, research shows that people often struggle to phrase prompts when using AI conversational agents to produce useful answers, and they often have trouble telling whether a prompt was effective~\cite{Zamfirescu2023FailPrompt, Jiang2022}. When answers are vague or incomplete, users sometimes try to “debug” the system by chaining or unit-testing prompts to see where things went wrong~\cite{Wu2022}. To reduce this effort, some systems have removed explicit prompting by giving users ready-made, high-level requests~\cite{Nam2024}. 
\colortext{For example, rather than requiring users to craft prompts from scratch, these systems provide pre-formulated templates (such as, summarize this document, explain this clearly) or suggested follow-up prompts tailored to the task at hand~\cite{Yuetian2025}.} 
Additionally, people can misjudge how much to rely on the AI; those who are overconfident in their own ability may ignore suggestions, while others may lean on the system too heavily~\cite{He2023}. 
\colortext{For example, when using AI for writing, people spend considerable time reading through AI-generated suggestions and deciding whether to accept, modify, or reject them, a process that requires constant judgment about the quality and appropriateness of each suggestion~\cite{Vaithilingam2022, sarkar2022like}. This creates a continuous need to monitor both the AI's outputs and one's own confidence in evaluating them.}
\colortext{Prior work shows that certain interface features, such as confidence cues, uncertainty visualization, or offering multiple alternatives, can help people monitor and review AI output more effectively~\cite{weisz2021perfection, Zamfirescu2023FailPrompt}.}
Recent work has looked at these challenges as metacognitive demands, the need for extensive monitoring and control for a task, and suggested design strategies that can either scaffold or reduce them~\cite{Tankelevitch2024}.
\colortext{These strategies include making outputs more transparent (for example, showing which sources informed the response or indicating confidence levels for different pieces of information) and allowing users to customize how information is presented (for example, adjusting response length, level of detail, or format preferences)~\cite{Tankelevitch2024}.} 
However, there is limited empirical evidence on how people engage in these processes when interacting with conversational agents in practice, particularly in domains such as health information seeking.

\subsection{Systems and Strategies Supporting Health Information Seeking with AI Conversational Agents}

In recent years, researchers have taken substantial steps to improve how conversational agents could support health information seeking. 
One line of work has focused on training these agents with medical data to better reflect domain knowledge and perform more reliably on health-related tasks.
For example, Med-PaLM and its successor Med-PaLM 2 have been trained on curated clinical datasets and evaluated them against medical licensing benchmarks, showing high performance in question answering and reasoning within medical domains~\cite{Singhal2025}. 
PMC-LLaMA~\cite{Wu2024PMC} and HuatuoGPT~\cite{zhang-etal-2023-huatuogpt} were built on general-purpose models and further trained with medical literature and patient–provider dialogues to increase factual accuracy and domain relevance. Others, such as Personal Health Large Language Model (PH-LLM), a Gemini-based model fine-tuned for personal health applications from wearable data, have demonstrated expert-level performance on case studies and exams~\cite{cosentino2024towards}.
In addition, a complementary line of work has focused on developing systems and interfaces that scaffold people's engagement with conversational agents for health information seeking. KnowNET, for instance, enhances agent responses by integrating external knowledge graphs, helping individuals explore medical topics through structured visualizations~\cite{Yan2025KNowNEt}. 
Similarly, DiscipLink scaffolds early-stage information seeking by generating tailored, domain-aware questions that prompt individual to clarify or expand their queries~\cite{Zheng2024DiscipLink}. 

While recent work has improved both the accuracy of AI conversational agents' outputs and the design of their interfaces, far less attention has been given to the metacognitive demands these systems place on people~\cite{Zamfirescu2023FailPrompt}. 
This burden is especially concerning in sensitive domains such as health, where the consequences of misunderstanding or misjudging information can be serious~\cite{Church2023}. If people do not question the validity of a response or reformulate a prompt, they may unknowingly accept flawed or incomplete information.
However, there is limited empirical research on how people experience and respond to these demands, particularly in the context of health information seeking. Our study takes steps to address this gap by exploring how individuals experience and respond to metacognitive demands when using AI conversational agents for health-related questions.

\section{Methods}

\begin{figure*}[t]
  \centering
  \includegraphics[width=0.9\linewidth]{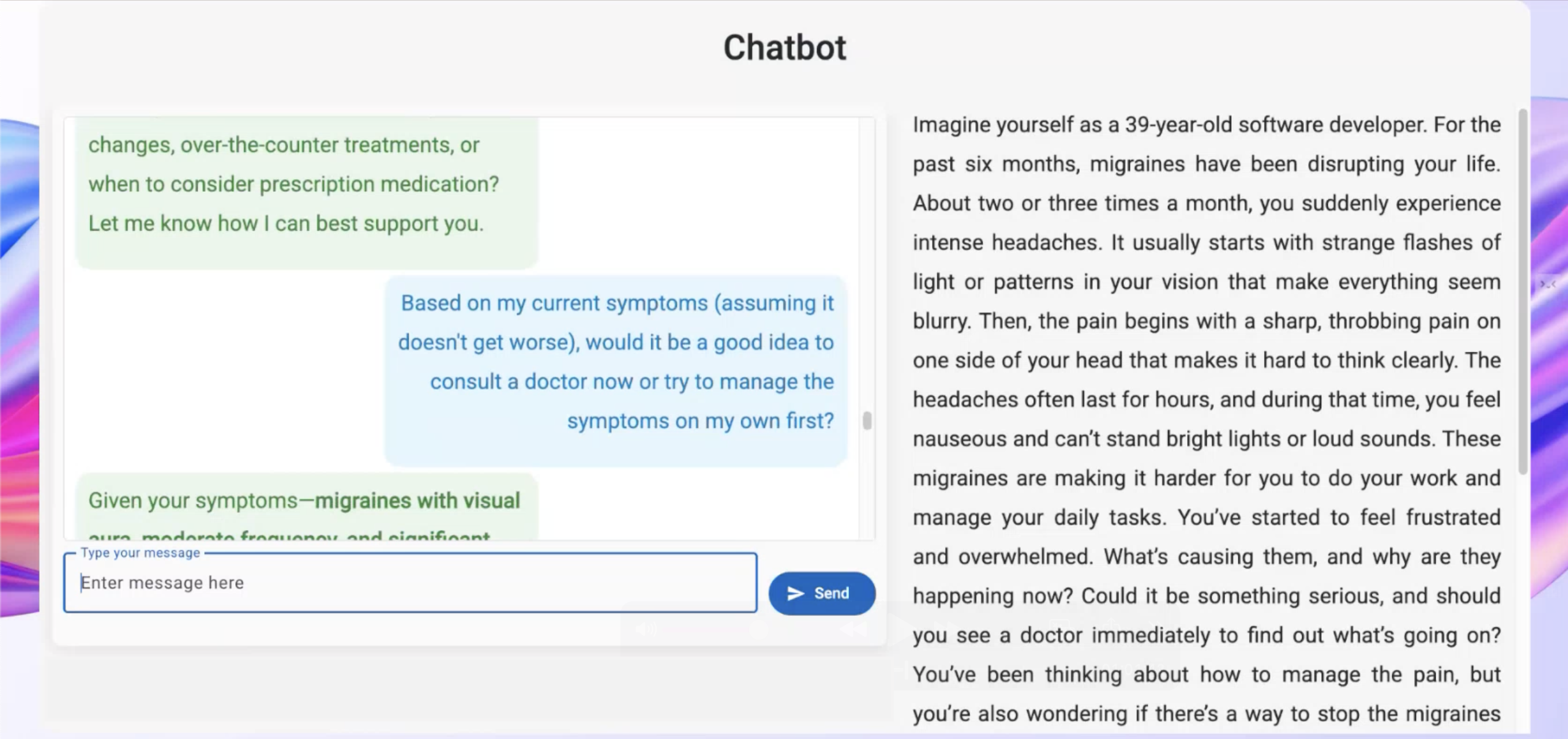}
  \caption{Screenshot of the AI conversational agent interface we developed to study the metacognitive demands encountered while gathering health information. This image shows an interaction from one of our participants during the study.}
  \Description{Figure 1: Screenshot of the AI conversational agent interface we developed to study the metacognitive demands individuals encounter while gathering health information. At the top of the interface, a header labeled ‘Chatbot’ is displayed. On the right side, the participant’s assigned health scenarios are listed. The main panel shows the chat window, where the participant has typed a question and the agent has provided a response. The interface also includes input controls below the chat for submitting new messages.}
  \label{fig:interface}
\end{figure*}

We aimed to understand the metacognitive demands people face and the strategies they use to cope with these demands when using off-the-shelf AI conversational agents for seeking health information.
We conducted a think-aloud study in which participants gathered health information using a conversational agent interface that we developed with the ChatGPT-4o API in an Angular environment (See Figure~\ref{fig:interface}). We built our conversational agent interface to resemble current off-the-shelf AI conversational agent interfaces. \colortext{We chose to develop our own interface rather than directly using existing platforms such as OpenAI's ChatGPT to maintain a consistent interface and model configuration throughout the study, as commercial off-the-shelf conversational AI systems regularly update their interface and underlying models. Building our own interface allowed us to securely store complete conversation logs within our research environment in accordance with our institutional ethics protocol, ensuring that all study data remained on our university's servers. Additionally, we were able to present the scenario text and the agent side by side on a single page, helping participants easily reference the scenario while formulating their queries.}

We drafted six health information–seeking scenarios for interpreting symptoms, identifying possible causes, and exploring treatment options for commonly encountered health concerns: allergies, insomnia, digestive issues, migraines, diabetes, and high blood pressure, and verified them with our clinician collaborator. 
We randomly assigned two scenarios to each participant and presented them one at a time. For each scenario, we asked them to imagine themselves in those situations and use the agent to gather relevant information, and talk about their process out loud. \colortext{This study was approved by Carleton University's ethics review board, and all participants provided written informed consent.}

\subsection{Health Information Seeking Scenarios Development}

We developed simulated health information–seeking scenarios that could serve as a stepping stone for exploring health information–seeking processes with conversational AI agents. 
To create these scenarios, we collaborated with an internal medicine physician, who advised on clinical plausibility, validated the medical accuracy of symptom descriptions and treatment options, and helped refine the scenarios to reflect situations a layperson could understand. 
Together, we developed six scenarios (See Appendix) that captured variation across health condition type (acute vs.\ chronic), time sensitivity (critical, urgent, standard, non-emergency), and potential for harm (high risk, rising risk, low risk).

\subsection{Participant Sampling and Recruitment}
To determine the number of scenarios for each participant, we conducted a pilot study with six human-computer interaction and AI researchers, which showed that two scenarios could be completed within a one-hour session while still giving participants enough time to acclimate to the interface without introducing fatigue. Thus, we assigned two simulated health information–seeking scenarios to each participant. 

Our goal in this study was to ensure balanced coverage across the six scenarios, with at least two participants engaging with each. 
Because scenarios were presented in pairs (6 choose 2 = 15 possible pairs), achieving full pairwise coverage required a minimum of 15 participants. 
\colortext{To capture a wide range of perspectives and account for the diverse characteristics of individuals in health information–seeking behavior, participants were recruited using a purposive maximal variation sampling method~\cite{palinkas2015purposeful}. 
This approach aims to identify common themes across heterogeneous samples, thereby providing a comprehensive understanding of how people interact with conversational agents for health information and the metacognitive demands they encounter. 
To achieve maximal variation, we used a priori recruitment quotas (See Appendix) targeting participants with diverse characteristics, guided by prior literature on health information–seeking behavior, including age~\cite{Fastrich2024Age}, gender~\cite{Rowley2017Gender}, education~\cite{Dutta2009education}, and frequency of using conversational agents (for example, ChatGPT, Claude, Google Gemini)~\cite{TechUse}.}
Inclusion criteria were adults aged 18 years or older who could speak and write in English and had at least a high school level education. \colortext{This ensured that participants could comfortably read and interpret the scenarios, understand the health information presented to them, and express their information needs clearly when formulating prompts.} Individuals unable to provide independent consent were excluded.
Participants were recruited through Prolific, Slack channels, a university research study Facebook page, email outreach to community groups (e.g., local library groups, community learning clubs, and neighborhood associations), and snowball sampling. \colortext{We used these multiple recruitment sources to meet our maximal variation sampling quotas and to reach a diverse set of participants. For instance, Prolific and Slack enabled us to reach younger adults and individuals who regularly use AI tools, whereas community groups and local libraries helped us connect with older adults and people with limited prior exposure to conversational agents. Social media outreach attracted participants with varied educational backgrounds, and snowball sampling extended our reach to individuals who may not be active on online platforms. }No contact information was collected about referred individuals; interested participants contacted the research team directly. \colortext{The eventual sample included participants from all recruitment sources, which ensured access to individuals who might not be present on a single platform and increased the diversity of experiences with conversational agents.} Each participant received \$20 compensation for their time.

\subsection{Data collection}
All study sessions were conducted virtually via Microsoft Teams, with both audio and screen activity recorded. Each session lasted approximately 60 minutes.
At the start of each session, participants provided informed consent and completed a pre-study questionnaire. This included demographic information such as age, gender, highest degree earned, and occupation. Additionally, we asked participants about their AI usage (i.e., frequency of AI use) and health information–seeking habits, including tools previously used for health information seeking, frequency of health searches, and self-rated confidence in finding health information on a 1–10 scale. \colortext{We also asked participants to complete a shortened, 19-item version of the Metacognitive Awareness Inventory (MAI)~\cite{Harrison2018} to gather their self-reported metacognitive awareness in the context of information seeking.}
For each scenario, participants were asked to imagine themselves in the described situation and use the chatbot to gather relevant health information (for example, understanding possible causes, considering treatment options, or deciding when to see a doctor). \colortext{Although the scenarios referenced common medical conditions that some participants might have experienced, we implemented safeguards to ensure the study posed no risk. All scenarios were presented as fictional, and participants were explicitly instructed that the conversational agent’s responses should not be interpreted as clinical guidance. The researcher monitored each session and intervened if participants attempted to interpret or apply any information as personal health advice.} A think-aloud protocol was used throughout, with the interviewer prompting participants to elaborate when clarification was needed. Participants were encouraged to explain how they formulated queries and interpreted responses. \colortext{No specific strategies for interacting with the agent were suggested, ensuring that participants’ actions reflected their natural approaches to interacting with the agent rather than being primed.}
We chose the think-aloud method because it allowed participants to verbalize their thought processes in real time while interacting with the agent to gather information.
This approach enabled us to capture how participants created prompts, used strategies, judged the agent’s responses, made decisions, and, when they faced uncertainty or confusion, how they dealt with it.

\subsection{Data analysis}
We used Microsoft Teams’ automatic transcription feature to generate transcripts after each think-aloud session. \colortext{To ensure accuracy, both coders reviewed the transcripts while simultaneously watching the session recordings, correcting any errors or omissions introduced by the automatic transcription.}
We then analyzed the transcripts data using NVivo and Microsoft Excel. 
Our analysis consisted of two major rounds. 
In the first round, two coders (the first and second authors) independently coded two transcripts, assigning descriptive labels (e.g., uncertainty in symptom description, unsure how to evaluate the response) to capture the ideas or concepts present in the data. Coders also reviewed the video recordings to capture details that were not always evident in the transcripts. 
These included participants retyping or reformulating prompts, pausing to reread responses, and scrolling back to earlier answers. 
Coders then compared their results, discussed discrepancies with the principal investigator, and developed a codebook (See Appendix) through iterative review of codes until full agreement was reached, which was then used to code the subsequent think-aloud sessions. In the second round, the same two coders independently organized the agreed-upon codes into the steps of the GenAI–human workflow~\cite{Tankelevitch2024}, \colortext{mapping each code, that is, the metacognitive demands and strategies, to a specific stage in the workflow. These stages included prompt formulation, evaluation, prompt iteration, and understanding and adapting workflows, which also resemble phases of health information-seeking behavior (searching, appraising, and applying information).  This workflow allowed us to capture how participants’ demands and strategies manifested at each step of their interaction with the AI conversational agent.}
\colortext{Consistency between coders was ensured through independent coding by two researchers, iterative discussion, and resolution of discrepancies with the principal investigator. However, we did not calculate inter-coder reliability mathematically. Thematic saturation was assessed throughout coding using a saturation table that documented the themes and codes identified in each interview, continuing until no new themes emerged.}

\section{Results}
A total of 15 individuals participated in our study. \colortext{All the recruitment quotas were met, resulting in a patient sample with diverse characteristics (see Table~\ref{tab:participants}).
We recruited 4 participants aged 18–29 (27\%), 6 aged 30–64 (40\%), and 5 aged 65+ (33\%), exceeding our >20\% target for each age group. We achieved gender diversity with 7 men (47\%) and 7 women (47\%), surpassing the >40\% threshold for each gender. We secured educational diversity by recruiting 4 participants with some college or less (27\%), 7 with bachelor's degrees (47\%), and 4 with graduate degrees (27\%), all above our >20\% target. Finally, we fulfilled AI conversational agent usage quotas with 4 never users (27\%), 6 occasional users (40\%), and 5 frequent users (33\%), each exceeding >20\%.}
Our analysis examined how participants sought health information using an off-the-shelf AI conversational agent.
\colortext{We analyzed participants' shortened MAI questionnaire responses and their think-aloud transcripts. Participants demonstrated high self-reported metacognitive awareness, with a mean MAI score of 16 (SD = 2.56) on a scale of 0-19 (see Appendix for participants' shortened MAI questionnaire responses). However, when participants verbalized their thought processes during the think-aloud sessions, they described substantial metacognitive demands, for example, the planning, monitoring, and evaluation required to interact with the agent, the demands that were not apparent from their MAI scores.
Thus, we focused on these think-aloud transcripts to identify the metacognitive demands participants encountered and the strategies they used to manage them.}
We organized our findings using the steps people took when interacting with the AI conversational agent—these steps have previously been described as prompt formulation, iteration, evaluation, and understanding and adapting workflows~\cite{Tankelevitch2024}. Within each step, we describe the metacognitive demands participants encountered and the strategies they used to cope with them. \colortext{Thematic saturation was reached after analyzing 12 interviews (See Appendix)~\cite{Saunders2018}.}

\begin{table*}[t]
\centering
\caption{Participant Demographics and Health Information Seeking Background.}
\label{tab:participants}
\small
\begin{tabular}{@{}p{0.3cm}p{0.4cm}p{1.4cm}p{1.3cm}p{1.9cm}
>{\raggedright\arraybackslash}p{2.1cm}
>{\raggedright\arraybackslash}p{3.4cm}
>{\raggedright\arraybackslash}p{1.8cm}
>{\raggedright\arraybackslash}p{2.3cm}@{}}
\toprule
\multicolumn{5}{c}{\textbf{Demographics \& Background}} &
\multicolumn{1}{c}{\textbf{AI Usage*}}
& \multicolumn{3}{c}{\textbf{Health Information Seeking Behavior}} \\
\cmidrule(lr){1-5} \cmidrule(lr){6-6} \cmidrule(lr){7-9}
\textbf{ID} & \textbf{Age} & \textbf{Gender} & \textbf{Education} & \textbf{Occupation} & 
\textbf{How often do you use AI conversational agents in general?} & 
\textbf{What tools have you used to find health information? } & 
\textbf{How often do you look for health information?} & 
\textbf{On a scale of 1–10, how confident are you in finding health information?} \\
\midrule
P1  & 28 & Woman      & Graduate     & Professor   & Most of time & Google, AI            & Monthly    & 6/10 \\
P2  & 21 & Man        & Highschool  & Student           & Always       & Google, AI, Books     & Monthly    & 8/10 \\
P3  & 25 & Woman      & Graduate     & PhD Student       & Most of time & Google, AI            & Monthly    & 6/10 \\
P4  & 33 & Woman      & Graduate     & Student           & Never        & Google                & Quarterly  & 5/10 \\
P5  & 26 & Man        & Bachelors    & Masters Student   & Sometimes    & Google, AI, WebMD     & Weekly     & 9/10 \\
P6  & 30 & Woman      & Bachelors    & Fitness Trainer   & Never        & Google                & Monthly    & 7/10 \\
P7  & 41 & Non-binary & Bachelors    & Administrative    & Most of time & Google, AI, WebMD, YouTube & Weekly & 7/10 \\
P8  & 70 & Man        & College & Retired           & Rarely       & Google                & Monthly    & 8/10 \\
P9  & 68 & Man        & Graduate     & Realtor           & Sometimes    & Google, AI            & Multi/week & 7/10 \\
P10 & 68 & Man        & Bachelors    & Caregiver         & Sometimes    & Google, AI            & Multi/week & 1/10 \\
P11 & 32 & Man        & College & Hairdresser       & Never        & Google                & Weekly     & 6/10 \\
P12 & 51 & Woman      & Bachelors    & Manager           & Sometimes    & Google                & Monthly    & 7/10 \\
P13 & 26 & Woman      & Bachelors    & Auditor           & Never        & Google                & Weekly     & 7/10 \\
P14 & 70 & Man        & College & Retired           & Most of time & Google, AI           & Monthly    & 9/10 \\
P15 & 78 & Woman      & Bachelors    & Retired           & Rarely       & Google                & Multi/week & 8/10 \\
\midrule
\multicolumn{9}{l}{\small \textit{*Referred to the use of any AI conversational agents (e.g., ChatGPT, Claude, Gemini)}} \\
\bottomrule
\end{tabular}
\end{table*}

\subsection{Prompt formulation: \emph{self-awareness, task decomposition, and well-adjusted confidence}}
Prompt formulation refers to the process of phrasing one’s query to the conversational agent throughout the interaction.
In our study, most participants (9/15) had difficulties with how to begin the conversation and what details to provide in their prompts. 
After reading the health scenarios and starting to chat with the agent to find relevant information, most participants (8/15) said that the blank input box with no cues left them uncertain about how to frame their prompt and what information to disclose. As P14 asked, \emph{``What should I type\ldots should I start with some of my symptoms or give my entire medical history?''} \colortext{This uncertainty persisted regardless of participants' prior experience with AI conversational agents. P14, who used AI most of the time, expressed the same confusion as P4, P6, and P11, who had never used AI agents before.}
Participants expected to see some form of health-specific guidance in the interface, for example, template prompts showing how to phrase information, what and when to disclose, but instead had to work out a starting point on their own. 
Participants believed that to receive tailored and personalized responses, they had to share personal details with the agent. Some participants (5/15) were willing to share personal details, but before doing so, they preferred to test the agent's trustworthiness. As P1 explained, \emph{``I don't want to spill everything out in the first [prompt] because I wanted to know basically how well is this [agent]''}. To test the system, participants staged their opening moves with \emph{``trial questions''} (P8)---for example, \emph{``what are migraines?''} (P13) or \emph{``what are all the possible causes of insomnia?''} (P9). These probes were not meant to learn new information but to check whether the agent's answers matched what they already knew. If responses strayed from expectations, participants did not disclose further information. However, not all participants chose to test the conversational agent before sharing. Some (4/15) felt it was safer to avoid disclosing personal details altogether. 
Instead of describing their own symptoms, those participants asked only broad, impersonal questions---such as \emph{``what are common causes of stomach pain?''} (P10) or \emph{``what are healthy sleep habits?''} (P12). This strategy let them stay in control of what they revealed, but it also meant giving up the possibility of personalized answers.
In this process, participants had to reflect on what they wanted from the agent, deciding which details were essential, weighing the risks of sharing the details, and choosing whether to hold back (i.e, \textit{\textbf{self-awareness of health goals and risks of information disclosure}}). 

Expressing bodily sensations in a way that the AI conversational agent could interpret them as intended was challenging for most of the participants (11/15). 
In a clinical setting, patients could rely on their clinicians to interpret their words, but while using the agent, they had to rely solely on their own words.
This created doubt about whether their descriptions were adequate.  As P4 explained, \emph{``for something like a feeling in the throat, I might just say uncomfortable or cannot swallow \ldots but with the agent, I had to think of all the other descriptive words.''} \colortext{Especially, all four participants who had never used AI conversational agents before faced the difficulty of formulating prompts to express their symptoms clearly, as well as deciding whether their prompts were accurate enough.}
Others worried about being misunderstood entirely: \emph{``I'm not sure it [agent] can understand what I'm trying to say.''} To cope with this, participants (6/15) often typed all the relevant symptoms they could think of into one long prompt. 
They reasoned that by including as many details as possible, at least some words might be picked up and lead to an accurate response. However, participants were uncertain whether the agent picked up all the details, because in its responses, they noticed it [agent] emphasized only a few points while overlooking others, \emph{``Why didn't it mention the others [symptoms]\ldots did it pick?''} (P11). 
These difficulties created a demand to structure their symptoms and break down their prompts into smaller texts (i.e., \textit{\textbf{task decomposition of symptoms and health queries}}).
To manage this demand, some (5/15) entered their input step by step across multiple turns, beginning with symptoms they could phrase clearly and introducing vaguer sensations later. Others (6/15) chose to lead with the detail they personally considered most serious, believing this would guide the agent's focus. As P12 explained, \emph{``I would start with the most concerning thing\ldots and then wait and see what questions it asks me.''}

Beginning the first prompt with the most serious symptoms made participants anxious (6/15) that they might neglect to communicate other key information: \emph{``I'm just scared of missing out some details, what if I miss something important and that shifts the entire response, I have to be sure since it's my health''} (P1). Participants had to carefully think about each word before pressing enter so they would not miss anything. They had to evaluate whether their wording was \emph{``good enough''} (P11) to explain their symptoms without misleading the conversational agent, and to continuously adjust their confidence in the completeness of those descriptions (i.e., \textit{\textbf{well-adjusted confidence in symptom description}}). As P10 explained, \emph{``I know if my question is misunderstood, so I will end up in a loop of rephrasing.''} 
In order to avoid overlooking important details, participants (6/15) looked for ways to strengthen their confidence in how they phrased their symptoms. Some (4/15) participants drew on strategies from their past clinic visits. They tried to recall the types of questions clinicians usually asked them, when a symptom began, how often it appeared, and how severe it felt, and used these as checklists for what to include. As one explained, \emph{``first to describe all my symptoms, then when and maybe how severe, that's what my doctor usually focuses on''} (P9). \colortext{Notably, older adults, four of the five participants aged over 65, used this approach, relying on their prior clinical notes to structure their prompts and guide what information to share.} A few participants (3/15) tried to build confidence in their wording by checking or borrowing terms from trusted sources before entering them into the conversational agent. They believed that if their phrasing matched what they found on their trusted sources, it would make the agent’s response more accurate. As P15 explained, \emph{“I’m going to Google search and see if it’s right.”} A few participants built their confidence in their wording (3/15) by describing their symptoms in plain, everyday language that reflected how they would normally talk about what they felt---for example, \emph{``burning in between the chest and a bit above the stomach after meals.''} They told us this was the easiest way to start, but they kept in mind that they could later switch to more precise or medical terms if the agent's replies seemed too broad.

\subsection{Evaluation: \emph{well-adjusted confidence}}
We refer to evaluation as the act of assessing the responses provided by the conversational agent for each prompt throughout the interaction.
When participants received responses from the agent for their questions, at first glance, the agent’s replies often looked comprehensive and complete, but participants struggled to judge what was relevant and trustworthy.
One challenge was the sheer amount of information; participants (7/15) noted that it sometimes provided more information than they had expected. For example, after a question about the causes of a symptom, the agent provided not only the causes but also the treatments and additional health warnings. P9 explained, \emph{``I asked for causes, it also listed what should I do, when should I see a doctor and more than what I asked.''} 
While the responses appeared thorough, participants often could not find the specific details they wanted, as these were buried within long lists, leaving them unsure whether their original question had been addressed. One participant said, \emph{``I’m not too confident now, I don’t know if mine [prompt] worked actually''} (P7). 
When responses came as long blocks of text without indicating which points were most important, many participants (8/15) felt unsure about how to interpret them. Some worried whether to treat all details equally or to focus only on certain ones. 
As P12 explained, \emph{``I could start with the first cause and then convince myself that that’s what it is.''} 
The overly confident tone of the responses also influenced participants' desire to evaluate them, \emph{``It sounds so sure, but I don’t know if it’s right''} (P11). \colortext{This concern was common among all participants who had never used conversational agents before. They expressed uncertainty about how the agent generated its answers and where the information came from.}
Because the system gave no clear signals about what to trust or how to weigh the information, participants had to do this work themselves. 
This created a constant demand on participants' minds to be confident enough to use the agent’s responses when making decisions about their health, but doubtful enough to stay alert to misleading or incorrect information (i.e. \textbf{\textit{well-adjusted confidence in weighing trustworthiness of health responses}}). 
To manage these long and dense responses, some participants (5/15) focused on bolded headings or highlighted phrases. P10 explained, \emph{``I like the highlighted causes\ldots I don’t want to read all; just pick a few highlighted and read in detail if needed.''} Others (4/15) skipped the full response and jumped directly to the closing summary, saying it was easier to evaluate and remember a short version than the full text, \emph{``The shorter version is easy for me to check and remember''} (P1). A few (3/15) scanned the text for words like \emph{doctor} or \emph{blood test}, which they treated as cues for action.  For P12, seeing ``blood tests'' was decisive: \emph{``That to me might be a prompt\ldots I probably really should book an appointment with the doctor anyway.''} Some (3/15), \colortext{all of whom were older adults,} rewrote the agent’s replies into their own notes, saying that putting the text in their own words helped them see what information they had already gathered and what was still missing. These strategies show how participants actively worked to steady their confidence when facing overwhelming replies. 

Another way participants (6/15) approached evaluation of responses was by checking whether the agent had correctly understood their prompt before reading the entire response. They often looked at the first few lines of the reply, and if the agent restated their question in a way that acknowledged the details they had given, this boosted their confidence, as it suggested the rest of the reply was likely to stay on track. P1 explained that summaries at the beginning of a response \emph{``made it clear it understood me.''} 
In contrast, when the agent replaced their words with unfamiliar medical terms or terms left unexplained, some participants (4/15) lost confidence in their evaluation because they assumed the reply no longer matched their original prompt. \colortext{Particularly, all four of these participants were older adults, and they often lost confidence whenever they encountered medical terms they did not know in the agent’s response.} In those cases, they often rewrote the prompt to reset the conversation. As P10 said, \emph{``I know my question is misunderstood, so I will back up and rephrase it.''} 

Even when participants felt that the agent had understood their prompt, they struggled to judge the accuracy of its answers. Several (5/15) compared and cross-checked the agents' answers against information from familiar sites such as Mayo Clinic or WebMD: \emph{``This information matches up with what I’m seeing on Mayo Clinic, which I know is like a trusted health website''} (P3). When using an agent, participants noticed that the source of information was not often provided, which made the information feel opaque and harder to verify. Without cues about where the content came from, participants said they had to fall back on their own judgment to decide whether to trust it.
A few (3/15) asked the agent for references, hoping citations would allow them to verify accuracy. But this was not always useful as the sources were sometimes too technical (e.g., research papers) or unfamiliar, which undermined rather than boosted their confidence. As P12 explained, \emph{``I’m not able to understand these references\ldots So I’m not 100\% convinced it’s accurate.''}

\subsection{Prompt Iteration: \emph{well-adjusted confidence and metacognitive flexibility}}
Prompt iteration refers to the process of refining or rephrasing one’s query to the agent, based on the previous response it produced.
After receiving the responses for prompts, some participants (9/15) felt that the responses sometimes did not fully answer their prompts (questions).
To make the response more relevant to their prompt, participants felt unsure of how to reword their prompt. 
Without clear guidance from the agent about what part of their input had been misunderstood or what details were missing, participants (7/15) had to rely on their own judgment and experiment to see whether a new wording would work better. 
To build this confidence, participants (8/15) tried out different ways of wording the same symptom until they felt certain that the agent had understood. For example, with dizziness, P14 clarified it as \emph{``not spinning, more like faintness''} so it would not be mistaken for something else. For head-related discomfort, others described how they went through chains of near-synonyms---\emph{``heavy head $\rightarrow$ head pressure $\rightarrow$ tight band around my head''}. They said each new attempt was a way of testing whether their words would ``land'' properly with the agent and give them more assurance that the prompt was effective. A few participants (3/15), \colortext{all of whom had never used conversational agents before,} believed that using medical terms could improve the accuracy of replies, so they asked the agent to rephrase their own words into clinical terms and then reused that language in later turns. Reusing the agent’s phrasing made them feel more confident that their next attempt would produce a more accurate response.
The work of prompt iteration demanded that participants adjust their confidence in whether their initial description of symptoms was being understood by the agent as they meant it, and to know when another rephrasing was needed to reach that clarity (i.e., \textbf{\textit{well-adjusted confidence in clarifying symptoms}}). 

Some participants (5/15) explained that no matter how many times they reformulated their prompts, the replies kept circling back to the same place. \colortext{This included four participants who had never used conversational agents and one who used them only occasionally.}
When caught in the repeated loop of typing similar prompts, they did not know whether to persist with minor tweaks or to change their approach altogether. This created a demand for flexibly adjusting prompt strategy by deciding when to abandon endless rewording of illness episodes to avoid getting trapped in a circle (i.e, \textbf{\textit{metacognitive flexibility to break rephrasing illness episodes}}). To exercise this flexibility, some (6/15) offloaded the work onto the agent itself. Instead of continuing to guess the best phrasing, they asked the system to take the lead: \emph{``ask me all the questions you need to help with my health condition''} (P2). Yet when the agent responded with an overwhelming list of questions, participants adapted again, narrowing the request to \emph{``ask me one question at a time''} (P1). Others (4/15) used another strategy and asked the agent to list potential symptoms so they could simply confirm if they are experiencing them: \emph{``Check back with me... you should have other symptoms, such as vomit. And I can say, oh actually I have vomit or don’t''} (P4).

Sometimes, participants (7/15) were uncertain or could not clearly recall their health details---especially about the timing or frequency of symptoms. They noticed that leaving a blank for those questions caused the agent to ignore the issue entirely. To avoid this, they often supplied rough estimates instead. As one explained, they would write \emph{``maybe three times a week''} or \emph{``not sure, but last month''} (P14). Providing an approximate answer felt safer than silence, but it also created new worries. Some participants (4/15), \colortext{all of whom had never used conversational agents before,} questioned whether rough estimates such as \emph{``maybe''} would make their responses seem less credible to the agent, or lead it to underestimate the seriousness of their condition. Participants told us that what they wanted was explicit confirmation that their description was sufficient, but the agent did not offer this. As one explained, \emph{``I was expecting... for the chatbot [agent] to say you have told me these details so far, but you haven’t told me any of the specifics related to this''} (P10). Without that kind of feedback, participants’ concern about incompleteness persisted. Sometimes participants (6/15) realized midway through the conversation that they had left something out. When this happened, they tried to patch the gap by adding forgotten details later---\emph{``I should have mentioned it’s worse at night''} (P10)---and then asked if the new information changed the advice. Yet most found that once the agent had already locked onto an explanation, late additions rarely shifted its reasoning. This pushed some (4/15) to restart the interaction entirely, \emph{``Well, let’s kind of start over again... I don’t want this symptom to be missed''} (P9). These restarts were effortful and time-consuming, but participants thought they were the only reliable way to prevent early omissions from influencing the conversation throughout.

Most participants (10/15), \colortext{irrespective of their age, education level, or prior use of AI conversational agents,} faced the problem of getting long, dense replies that were hard to sift through. To address this, they changed how they instructed the agent to present its responses. They asked for clearer structures---short lists, quick summaries, or ordered explanations---believing that if the reply appeared in an organized way, it meant their prompt had been effective. P9, for instance, asked it to \emph{``list causes... from least serious to most serious,''} while P1 requested, \emph{``quickly summarize what I can do in the meantime.''} When the agent followed these instructions, participants (6/15) said it boosted their confidence that they had found the ``right way'' of asking. But in response to the next prompts, the agent soon slipped back into long, verbose paragraphs, \emph{``It’s too verbose... it went back to its usual way again''} (P1). In these situations, participants felt that they lost their sense of control and confidence in this strategy. This left many frustrated.

\subsection{Understanding and Adapting the Workflow: \emph{self-awareness, well-adjusted confidence, and metacognitive flexibility}}
We refer to understanding and adapting the workflow as the process of recognizing when and where the agent’s input should be considered within one’s health decision-making process, and adjusting one’s workflow to accommodate that role.
Some participants used the agent as a tool for gathering and organizing background knowledge (6/15), while others wondered whether its suggestions could---or should---play a role in decision-making (5/15). Because the agent’s responses to participants’ questions often came across as final answers, they were tempting to trust and sometimes began to be taken as definitive decisions. As P4 said, \emph{``It gives plausible answers… it seems true… but it does not… ultimately mean it’s right, right?’’} \colortext{All four participants who had never used AI conversational agents experienced this challenge,} explaining that without regularly stepping back to remind themselves what the agent was useful for---and what it was not---they risked over-reliance. These situations created a demand to be self-aware of the agent’s role and create boundaries for making health decisions (i.e., \textbf{\textit{self-awareness of the agent’s role in health decisions}}).
Participants generally described the agent as helpful for tasks such as listing symptoms or possible causes, but not for making health decisions. P10 explained, \emph{``I can use it for understanding, but not for my decisions. I almost feel like I will just trust this.’’} Similarly, P2 emphasized treating its replies as brainstorming material rather than conclusions: \emph{``I will not look for… a final answer but for the model’s various suggestions.’’} 
Some participants built boundaries for the agent; for example, they treated the causes it suggested as ideas to investigate further rather than as facts to accept immediately (7/15). P9 noted that while the agent provided \emph{``a good plan of this summary of what I should do next,’’} they would always \emph{``double check with other sources’’} before acting. Others created cut-off rules---deciding, for instance, to stop consulting the agent once it mentioned medications or tests (4/15). These strategies helped participants maintain their perspective when the agent’s tone was overly confident.  

Some participants pointed out that the agent occasionally included disclaimers such as \emph{``I’m not a doctor’’} (5/15). These reminders helped them draw a line between background information and medical advice. As P4 explained, \emph{``It understood that it provides general things. I would be surprised if it says you should listen to my healthcare advice for sure, like I’m an expert.’’} While some found the disclaimers helpful (3/15), others found them confusing because they appeared inconsistently---present in one reply but missing in another. When this happened, participants wondered whether the agent’s role had shifted from offering information to giving guidance, which left them less confident about how much weight to give the next piece of advice.  This uncertainty placed a constant demand on participants to adjust their confidence in safely delegating health information seeking to the agent (i.e, \textbf{\textit{well-adjusted confidence in relying on the agent without letting it drive health decisions}}). They valued the agent’s ability to structure and summarize information (9/15), but worried that leaning too heavily on it would impact their care decisions. As P9 reflected, \emph{``It gives me a good plan, and should I just go with it… it’s my health, do I have to think twice.’’} To manage this tension, some participants reinforced a clear division of roles: the agent could help organize, but responsibility for decisions stayed with them (6/15). As P2 explained, \emph{``I can use it for organizing what I know, but I still have to be the one to decide.’’}  

Even after participants had set boundaries for the agent’s role and clarified how they wanted to use it in their health information seeking, they occasionally found the system taking over the flow of the conversation (7/15). When they did not pause to recognize this shift, the interaction risked becoming more consuming than useful. This situation created a demand to adapt strategies by making decisions about whether to follow the agent’s lead or redirect the conversation to keep their health information seeking aligned with their goals (i.e., \textbf{\textit{metacognitive flexibility in adjusting workflow strategies in response to agent-led shifts}}). One common way this drift occurred was when participants (8/15) relied on the agent’s follow-up prompts to keep the conversation moving, rather than phrasing their own questions with intention. While these suggestions offered helpful directions at times, participants realized that following them without critical reflection could lead to detours into topics they had not planned to explore. \colortext{Participants faced this drift regardless of their prior experience with AI conversational agents. For example, P2, who regularly used AI conversational agents, and P13, who had never used them before, both faced the drift after following the agent’s suggested follow-up prompts without critical reflection.}  P13 reflected on this experience: \emph{``I was going to ask more about the treatment, but then it asked me about my lifestyle, and I just followed that instead.’’} To regain control, some described pausing the exchange to reset and reaffirm their priorities. One explained, \emph{``I had to stop and remind myself---I’m going to type: my priority was asking about treatment, so can we go back to that and not get distracted?’’}(P9).

\begin{figure*}[t]
  \centering
  \includegraphics[width=0.9\linewidth]{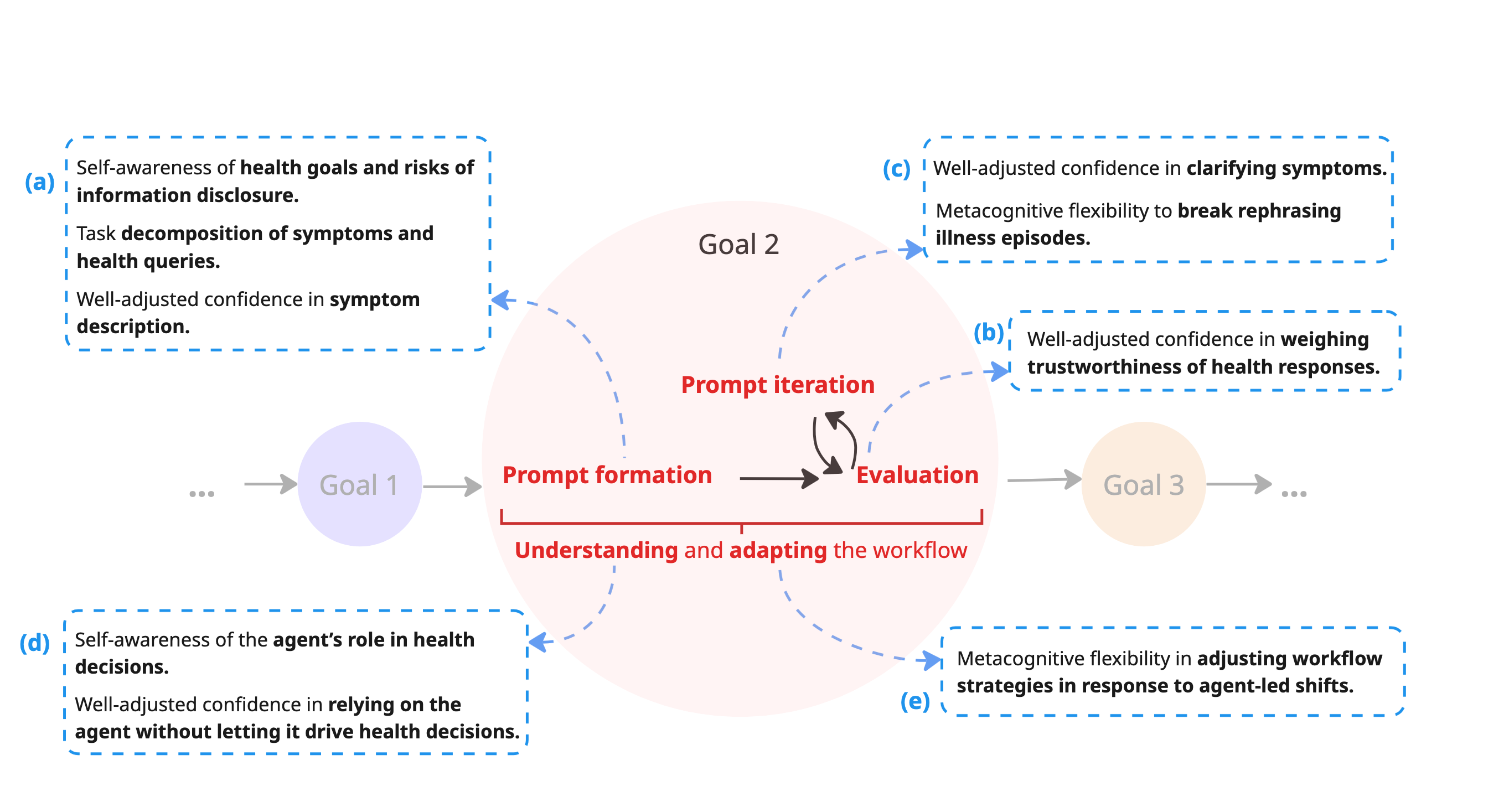}
  \caption{Summary of the metacognitive demands placed on people using AI conversational agents for health information seeking. Inspired by prior work on metacognitive demands \cite{Tankelevitch2024} and empirically observed in our study, the figure highlights the demands that were posed on people across health information seeking goals, zooming in on one goal as an illustration. Specifically, metacognition demands raised by AI agents during (a) prompt formulation, (b) evaluation, (c) prompt iteration, (d) understanding the workflow, and (e) adapting the workflow.}
  \Description{Figure 2: Screenshot of the GenAI Human Workflow for health information seeking. The figure shows five stages: (a) Prompt formulation, (b) Evaluation, (c) Prompt iteration, (d) Understanding, and (e) Adapting the workflow. Each stage includes descriptions of the metacognitive demands. Stage (a) includes self-awareness of health goals and disclosure risks, task decomposition of symptoms and questions, and well-adjusted confidence in symptom description. Stage (b) includes well-adjusted confidence in evaluating the trustworthiness of health responses. Stage (c) includes well-adjusted confidence in clarifying symptoms and metacognitive flexibility to break rephrasing illness episodes. Stage (d) includes self-awareness of the agent’s role in health decisions and well-adjusted confidence in relying on the agent without letting it drive health decisions. Stage (e) includes metacognitive flexibility in adjusting workflow strategies in response to agent-led shifts. The five stages form a loop and are repeated for each health goal (e.g., Goal A, Goal B, Goal C).}
  \label{fig:meta}
\end{figure*}
\section{Discussion}

\colortext{
Previous research has examined differences in information-seeking behavior across modalities~\cite{Kaiser2025AIvsGoogle}, such as conversational agents versus traditional web search, and highlighted the increasing use of AI conversational agents for health information~\cite{Wardle2025health}.
Thus, with the increased popularity of the use of AI conversational agents, the goal of our study was to explore how people use AI conversational agents for health information seeking, the metacognitive demands they encounter, and the strategies they adopt to cope with these demands. 
}
For analysis, we adapted the GenAI Human Workflow Framework~\cite{Tankelevitch2024} to the domain of health information seeking, using it as a lens to map the metacognitive demands people experience when seeking health information {(See Figure \ref{fig:meta})}. We found that when seeking health information using conversational agents, people often pursue several goals at once—such as making sense of symptoms, checking urgency, or preparing for a doctor’s visit—and for each goal, they must work through a sequence of steps. The process begins with \textbf{formulating a prompt}. This stage requires self-awareness of what the person is trying to achieve and caution about how much personal detail to share. It also involves structuring symptoms and breaking questions into smaller pieces while judging whether the symptom description feels complete enough.
After receiving a response, people move to \textbf{evaluating the response}. Here, they must weigh whether the agent’s advice is trustworthy, whether it has incorporated all of their reported symptoms, and whether it highlights urgent red-flag signs. Based on their evaluation of responses, \textbf{iterating on the prompt} becomes necessary. At this stage, people need to adjust their confidence in whether their initial description of symptoms was understood by the agent as intended, and recognize when another rephrasing is needed to reach clarity. At the same time, they must remain flexible enough to avoid falling into unproductive loops of rewording when the agent stays anchored on an earlier framing. Across these interactions, \textbf{understanding the agent’s role} in health decisions places demands on people to remain self-aware, set clear boundaries for decision-making, and adjust their confidence in relying on the agent without letting it dominate health decisions. Finally, while \textbf{adapting their workflow}, people also require metacognitive flexibility: deciding when to continue with the conversational agent, when to redirect the exchange to better match their goals, and when to disengage to seek professional care.

\subsection{Design Considerations}
Drawing on the metacognitive demands and coping strategies identified in our study, we propose five design considerations for developing future AI conversational agent interfaces tailored to health information seeking with affordances that guide individuals' actions. 
These considerations outline how agents could support people across the health information–seeking workflow: by scaffolding goal-setting, guiding the structuring and refinement of prompts, managing disclosure boundaries, showing the input details the agent has taken into account, and supporting reflection on when information is sufficient to act. 
We hope these considerations inspire researchers to move beyond current off-the-shelf systems and towards designing conversational agents that align with the health information seeking needs while reducing the metacognitive demands. Our design considerations are as follows:

During participants’ engagement with the conversational agent, they often struggled to keep track of their health information needs. While formulating prompts, even when they had a goal in mind, many were unsure how to communicate it with the agent. The agent frequently produced more information than they had asked for, which made people lose focus and drift into side topics. In health contexts, such drifting is particularly problematic because it can prevent people from asking the questions they came for—such as whether symptoms require medical attention—ultimately delaying or derailing care decisions. Yet current off-the-shelf AI conversational agents offer no cues to help people anchor or revisit their goals~\cite{Kim2024, Liang2024}, leaving them to manage this work of being self-aware of their goals entirely on their own. 
One way to address this challenge could be through a lightweight “mini-agent” embedded in the interface. At the start of a conversation, the mini-agent could prompt people to select their health information needs from commonly discussed health topics identified in prior literature~\cite{Martina2016}—for example, illness information, symptoms, prognosis, treatment, nutrition, side effects, or mental health. Once chosen, these categories could remain visible and editable throughout the chat, helping people stay oriented to their intentions. The agent could also constrain responses to the selected goals, ensuring focused replies rather than overwhelming detail. If the conversation drifted into unrelated areas, the mini-agent could gently check in—for example: \emph{“You chose symptoms and treatment as your goals for this conversation, but this question looks more like lifestyle. Would you like to add that?”}

\begin{list}{}{\leftmargin=1.5em}
\item \textbf{Design Consideration \#1:} Future conversational agents for health information seeking should include explicit goal-setting scaffolds that help people articulate their health information needs at the start, and track and stay aligned with those needs in the conversation. By supporting both the start and the ongoing flow of the interaction, such scaffolds could help people remain aware of their goals and avoid drifting away under the agent’s influence.
\end{list}

\noindent Breaking large queries/prompts into smaller parts is often recommended when using conversational agents, as the system can focus on one topic at a time and produce clearer answers~\cite{Ross2023}. 
This strategy works well in domains such as creative writing~\cite{Dang2023} or coding~\cite{Vaithilingam2022}, where tasks can be segmented. 
In health information seeking, however, decomposition appears to work unevenly. Some health information—for example, describing symptoms—needs to be provided together to preserve important context. For instance, describing migraines with visual disturbances can generate different responses from the description of migraines with headache alone. 
Receiving multiple possible causes for the symptoms could create health anxiety as participants imagined themselves in the worst-case scenario~\cite{wang2024there}.
Conversely, asking about causes, treatments, self-care, and when to seek medical attention all at once often resulted in overwhelming, lengthy responses that were difficult to digest and identify key points. Therefore, such questions should be broken down to avoid overloaded responses.
However, without guidance from the agent, participants were unsure how to approach this process and relied on trial and error.
In effect, they had to decompose the tasks themselves—deciding when to bundle information together and when to split it into smaller queries, which increased effort and created uncertainty about whether they had asked the questions in the ``right'' way. 
To address this challenge, one approach could be to support prompt structuring from the start of the interaction. Instead of relying solely on an open text box, the system might offer structured templates for symptom entry. 
Most of the recent research on prompt engineering focuses on text-based templates~\cite{Yuetian2025,  Zamfirescu2023FailPrompt}. 
However, many participants in our study reported difficulty articulating symptoms in words. 
We therefore suggest that AI conversational agents for health may need to move beyond text-only input templates and incorporate visual supports.
For example, an interactive body map\colortext{~\cite{Galve2020} }could be provided and allow people to select affected areas in the body map and add descriptors—such as sharp versus dull pain, or constant versus intermittent, using interaction techniques, such as sliders or toggles. This approach may help capture critical details in a single comprehensive prompt, reducing fragmented or incomplete symptom descriptions. 
After symptoms are captured, the system should support transition into adaptive questioning. Rather than leaving individuals to phrase broad, long queries, the agent should guide them toward focused questions tailored to their reported symptoms, for example: \emph{``Would you like to begin with possible causes?''} or \emph{``Do you want information on when to seek professional care?''}  

\begin{list}{}{\leftmargin=1.5em}
\item \textbf{Design Consideration \#2:} Future designs of AI conversational agent interfaces for health information seeking should support people in structuring their prompts by clarifying when complete information is needed (e.g., symptom descriptions) and when it is better to separate questions (e.g., follow-up queries). This approach can reduce trial and error, enabling more effective and focused interactions.
\end{list}

\noindent Participants were unsure whether to share personal and sensitive health details when interacting with conversational agents or rely on generic information. 
This tension arose from fears of oversharing that could lead to privacy breaches, unwanted exposure, or misuse of their data~\cite{Yin2025, Wester2024}. 
In healthcare settings, disclosures occur under strict confidentiality laws and protocols, which ensure that sensitive health information is handled securely and only for legitimate medical purposes~\cite{George2003}. 
In contrast, conversational agents lack such transparent safeguards, leaving individuals uncertain about what data may be retained, shared, or repurposed. 
This ambiguity raised concerns about unauthorized access, commercial exploitation, and accidental exposure of traumatic or stigmatizing information~\cite{patil2023sensitiveinformationdeletedllms}. 
As a result, participants had to remain self-aware of what to disclose, which often led them to withhold critical clinical details or abandon the use of the agent entirely. 
One way to support people in these situations is for the agent to provide short, lay-language onboarding statements that clarify the agent’s role and limitations. Echoing previous research on the importance of medical disclaimers~\cite{Lawrence2024}, we also suggest including clear statements such as \emph{``I am an AI information tool, not a medical professional''} or \emph{``I cannot diagnose, but I can provide general health information''}. 
People could also be informed that their inputs may be stored and used to improve the system unless they opt out. Our study further highlighted the importance of not disclosing personal health information to conversational agents.
Therefore, we recommend conversational agents to provide concrete examples of unsafe inputs (prompts) to individuals while in use. These examples could be personal identifiers (full names, addresses, phone numbers, emails, insurance IDs) or highly sensitive health experiences, including detailed trauma accounts, suicidal thoughts, or intimate disclosures that require professional, confidential care rather than AI guidance. 
Even with clear disclaimers, sometimes, people may still unintentionally share sensitive information. To mitigate this, we recommend embedding prompt sanitization—an automated check that reviews user inputs for risky content before they leave the device to the agent. Building on recent work~\cite{Zhou2025}, this could be achieved through a lightweight real-time detector that runs locally (e.g., as a browser extension) and scans text for patterns such as personal identifiers (names, phone numbers, addresses) or highly sensitive health details. When flagged, the system could not silently block the input; instead, it could pause transmission, highlight the risky content, and explain in plain language why sharing it could be unsafe. The agent could then suggest safer alternatives, so people remain in control of what to disclose. For example, if someone typed a full birth date, the detector could flag it and recommend sharing only an age range.  

\begin{list}{}{\leftmargin=1.5em}
\item \textbf{Design Consideration \#3:} Future health information–seeking AI conversational agents should safeguard sensitive health information disclosures by providing clear upfront guidance on what is safe to share and reinforcing these boundaries with real-time protective cues. These approaches can alleviate people’s worry around disclosure risks, help them understand the privacy protections the agent provides, and protect them from oversharing personal and sensitive information.
\end{list}

\noindent When explaining symptoms to conversational agents, participants frequently worried whether the system was incorporating all relevant details or focusing only on part of their input when generating responses. 
Echoing prior research~\cite{Liao_2020}, our findings show that this lack of clarity about how their input was used reduced participants’ confidence in the reliability of the health advice, since overlooking even a single symptom could shift how a condition was understood. 
Moreover, when participants added or corrected details later in the conversation—such as the appearance of additional or worsening symptoms—they observed that the new information was often ignored or inconsistently integrated once the agent had anchored on an early framing. 
This aligns with prior observations of conversational agents exhibiting anchoring effects, where initial framings disproportionately shape subsequent outputs~\cite{Santhanam2020, Nourani2021}. 
But in health contexts, the consequences can be severe: failing to update a case with new details may obscure red-flag signs or delay recognition of urgency. 
As a result, participants felt compelled to adapt by either continuing with the interaction, possibly correcting the misleading response trajectory or restarting the interaction altogether.
Current off-the-shelf agents provide little transparency about how they interpret symptom input~\cite{Bhatt2021}, leaving people to guess which details the system relied on and whether newly added information was considered. This lack of visibility makes it harder for individuals to trust the advice they receive and increases the risk of incomplete or unsafe health guidance and decisions.
One possible way to address these challenges would be to make the details the agent has taken into account visible.
For instance, a persistent side panel could summarize the agent’s current interpretation in lay language after each prompt (e.g., \emph{“You mentioned throat discomfort lasting two days with symptoms X and Y. Is that correct?”}). 
Individuals could then confirm or correct this understanding before a complete response is generated. 
The same panel could also update dynamically as new information is added mid-conversation, flagging changes and offering a simple “re-run'' with these details option. 
This approach could both reassure people that their input was understood correctly and allow late-arriving symptoms—especially urgent ones—to be incorporated without restarting the exchange.  

\begin{list}{}{\leftmargin=1.5em}
\item \textbf{Design Consideration \#4:} Future health AI conversational agents should provide input transparency by clearly showing how a person’s reported symptoms are being recorded and interpreted. A persistent, user-editable summary of the agent’s understanding could allow individuals to confirm that key health details are captured, ensure that new or worsening symptoms are incorporated, and reduce the risk of advice based on incomplete or outdated information.
\end{list}

\noindent People turn to AI conversational agents to gather information for health decisions, but it can be difficult for people to judge if they received the ``correct'' and ``enough'' details to make care decisions~\cite{Ayers2023}. Long responses bury key points within paragraphs, with no easy way to review, evaluate, or reflect. As a result, participants in our study had to both adjust their confidence in evaluating reliability and remain self-aware of whether they had gathered enough to act. The current off-the-shelf interfaces offer little to no support for either task, leaving individuals to decide on their own whether the information is reliable and sufficient to proceed. Prior work suggests that structured formats help people scan and prioritize information more effectively~\cite{Ma2024}. In our study, participants also expressed a preference for structured presentations, including contrastive or visual formats that help weigh options without bias toward the first detail seen. \colortext{Prior work also has shown that visual formats, such as pictographs, icon arrays, and comparative tables, improve comprehension and reduce misinterpretation of health information~\cite{Kandel2025GraphicalPO}.}
Based on our results, we recommend that health AI conversational agents provide visuals, such as tables to compare causes and symptoms, mind maps to show how symptoms relate, or timelines to highlight \emph{“what to do now”} versus \emph{“what to do later.”} Building on this, we also suggest presenting health information contrastively—for example, separating urgent symptoms from those safe to monitor, or distinguishing more likely from less likely causes—rather than burying everything in text. \colortext{When AI conversational agents are developing visuals for presenting health information, we also recommend using designs that minimize cognitive load. Studies suggest that using metaphorical or skeuomorphic design elements can facilitate sensemaking by grounding abstract concepts in familiar spatial patterns and everyday perceptual experiences~\cite{Stroke2024}. For example, instead of presenting symptom severity as numerical scales or abstract categories, agents could use visual metaphors such as traffic light indicators for urgency (green for monitor at home, yellow for call doctor soon, red for seek immediate care). }   
To further support people in determining whether they have gathered enough information to make informed health decisions, we suggest offering a reflection space adjacent to the chat interface. In health contexts, this is especially critical because missing or misinterpreting a single symptom can change whether a situation appears routine or requires urgent care. Drawing inspiration from prior whiteboard-style canvases~\cite{Xu2024}, a reflection space could allow users to save relevant excerpts from the agent’s responses or add personal notes, helping them track what has been covered across categories such as symptoms, possible causes, treatment options, and next steps. 
Building on our findings, we recommend enhancing the note-taking space with prompts tailored for health decision-making. For example: \emph{“Does this cover all the symptoms you described?”} or \emph{“Are you ready to decide whether to monitor at home or seek care?”} Such prompts could encourage people to pause and assess the sufficiency of their actions, reducing the risk of prematurely relying on incomplete health advice or endlessly looping through more questions.

\begin{list}{}{\leftmargin=1.5em}
\item \textbf{Design Consideration \#5:} Future health AI conversational agents should go beyond text by using visuals such as tables, mind maps, or timelines to clarify symptom–cause–treatment relationships, present health information contrastively (e.g., urgent vs. non-urgent symptoms, more vs. less likely causes), and provide a reflection space with prompts that help individuals verify whether red-flag symptoms, possible causes, and next steps have been addressed. This approach can reduce overload, strengthen evaluation, and support safer health decisions.
\end{list}

\subsection{Limitations and Future Work}
To the best of our knowledge, this is the first study to examine metacognitive demands and strategies while using AI conversational agents for health information seeking. However, several limitations should be considered when interpreting our findings. 
First, although we recruited participants with diverse backgrounds using maximal variation sampling, the overall sample size of fifteen was small. \colortext{As with all qualitative studies, our goal was not statistical generalizability to wider populations, but to provide a deep understanding of the metacognitive demands people face when using conversational AI for health information seeking. Although our sample represented a range of educational backgrounds from high school to graduate degrees the majority (13 participants out of 15) reported moderate to high confidence (scores in the range of 5 to 9 out of 10) in finding health information. Consequently, the results may reflect perspectives of individuals who are relatively comfortable engaging with health information and are able to read, write, and create prompts. Individuals with lower education or less confidence in gathering health information might experience these metacognitive demands differently, for instance, they may face greater difficulty articulating their information needs in text, rely more on surface cues such as tone or empathy rather than evaluating content accuracy, or feel less confident in judging the reliability of generated responses. Thus, future work with larger samples can establish the generalizability of these findings and can examine how these experiences vary across health literacy levels, cultures, and age groups.}
Second, participants engaged with simulated health scenarios in one-hour, single-session interactions. 
While these scenarios served as a stepping stone for examining health information–seeking processes with conversational agents, they may not fully capture how people’s strategies evolve with repeated use or the urgency and emotional stakes that arise when seeking information for their own ongoing health concerns. \colortext{We also acknowledge that real health situations involve heightened anxiety, time pressure, and personal consequences that may amplify metacognitive demands.}
Future research should therefore explore the longitudinal use in relation to participants’ personal health needs, examining how metacognitive demands and coping strategies develop over time, influence actual care-seeking decisions, and impact the long-term sustainability of engagement.
Conversational AI technology is evolving rapidly. Our study was conducted using a system built on GPT-4o API, but the development of newer models may mitigate some of the challenges we observed or introduce new ones. 
Continued research will be necessary to track how these evolving capabilities reshape the metacognitive demands of health information seeking and the strategies people use in response.

\section{Conclusion}

People increasingly turn to AI conversational agents for health information seeking. 
To understand how metacognitive demands arise when people use off-the-shelf AI conversational agents for health information seeking, and what strategies they adopt to cope, we conducted a think-aloud study with 15 participants. We provided them with health scenarios, asked them to imagine themselves in that situation, and gather relevant health information using our AI conversational agent interface. 
We analyzed the think-aloud sessions and identified the metacognitive demands these systems impose and the strategies people employed in response. 
Drawing from these insights, we outlined design considerations for moving beyond off-the-shelf interfaces to reduce metacognitive demands and support better user experiences and affordances in health information seeking.

\begin{acks}
We thank our participants whose invaluable contributions made
this research possible. We are grateful to the reviewers for their helpful feedback. Additionally, we extend our appreciation to
the Natural Sciences and Engineering Research Council of Canada
(NSERC), and National Research Council Canada, for their
generous support.
\end{acks}

\bibliographystyle{ACM-Reference-Format}
\bibliography{ref}

\end{document}